\documentclass[12pt]{article}

\usepackage{latexsym,amsmath,amssymb,fancybox}

\usepackage[english]{babel}

\usepackage{color,epsfig}
\usepackage{graphicx}

\definecolor{navy}{rgb}{0.0,0.0,0.4}

\definecolor{rd}{rgb}{1,0,0}
\definecolor{or}{rgb}{1,.33,0}
\definecolor{pi}{rgb}{.66,.33,.33}
\definecolor{gn}{rgb}{0,.50,0}
\definecolor{be}{rgb}{0,0,.66}
\definecolor{ru}{rgb}{.66,0,.33}
\definecolor{vi}{rgb}{.33,0,.66}
\definecolor{gy}{rgb}{0,.33,.66}
\definecolor{ye}{rgb}{.66,.33,0}
\definecolor{bk}{rgb}{0,0,0}

\font\srm=cmr8

\def\thf{\baselineskip=\normalbaselineskip\multiply\baselineskip
by 7\divide\baselineskip by 6}

\def\fff{\baselineskip=\normalbaselineskip}

\def\spose#1{\hbox to 0pt{#1\hss}}
\def\lta{\mathrel{\spose{\lower 3pt\hbox
{$\mathchar"218$}}\raise 2.0pt\hbox{$\mathchar"13C$}}}  \def\gta{\mathrel
{\spose{\lower 3pt\hbox{$\mathchar"218$}}\raise 2.0pt\hbox{$\mathchar"13E$}}}

\def\Euro{\spose {\lower 2.5pt\hbox{${^{\bf =}}$}}{ C}}

\def\spose#1{\hbox to 0pt{#1\hss}}

\def\sqr#1#2{{\vcenter{\hrule height.4pt\hbox{\vrule width.8pt height#2pt
\kern#1pt\vrule width.8pt}\hrule height.4pt}}}

\begin{document}

\def\be{\begin{equation}}
\def\fe{\end{equation}}

\newcommand{\eqn}{\label}
\newcommand{\bel}{\begin{equation}\label}

\def\eqdef{\fff\ \vbox{\hbox{$_{_{\rm def}}$} \hbox{$=$} }\ \thf }

\def\ov{\overline}


\def\Lr{ {\color{rd} {L}} }

\def\calIr{ {\color{rd} {\cal I}} }
\def\Vr{ {\color{rd} {V}} }
\def\Ar{ {\color{rd} {A}} }

\def\Br{ {\color{rd} {B}} }
\def\Cr{ {\color{rd} {C}} }
\def\Dr{ {\color{rd} {D}} }

\def\Xr{ {\color{rd} {X}} }\def\Yr{ {\color{rd} {Y}} }
\def\Er{ {\color{rd} {E}} }
\def\Rr{ {\color{rd} {R}} }

\def\calMr{ {\color{rd} {\cal M}} }

\def\nablar{ {\color{rd} {\nabla}} }

\def\calXr{ {\color{rd} {\cal X}} }
\def\calUr{ {\color{rd} {\cal U}} }
\def\calVr{ {\color{rd} {\cal V}} }
\def\calPr{ {\color{rd} {\cal P}} }
\def\calBr{ {\color{rd} {\cal B}} }
\def\Phir{ {\color{rd} {\Phi}} }
\def\Gammar{ {\color{rd} {\Gamma}} }
\def\Thetar{ {\color{rd} {\Theta}} }
\def\grd{ {\color{rd} {g}} }
\def\srd{ {\color{rd} {s}} }

\def\dr{\spose {\raise 4.0pt \hbox{\color{rd}{\,\bf-}}} {\rm d}}


\def\Gru{ {\color{ru} {G}} }
\def\kru{ {\color{ru} {k}} }
\def\calAr{ {\color{ru} {\cal A}} }
\def\calGr{ {\color{ru} {\cal G}} }
\def\calPr{ {\color{ru} {\cal P}} }
\def\calCr{ {\color{ru} {\cal C}} }
\def\ConStruc{ {\color{ru} {\copyright}} }
\def\omegaru{ {\color{ru} \omega}}
\def\Omegaru{ {\color{ru} \Omega}}

\def\Dru{ {\color{ru} {D}} }
\def\aru{ {\color{ru} {a}} }
\def\Aru{ {\color{ru} {A}} }
\def\Fru{ {\color{ru} F} }
\def\amr{ {\color{ru}\bf{a}} }
\def\Amr{ {\color{ru}\bf{A}} }
\def\Fmr{ {\color{ru}\bf{F}} }
\def\wru{ {\color{ru} {\vert\!\!\vert\!\!\vert}} }

\def\gbe{{\color{be} g }}
\def\sbe{{\color{be} s }}
\def\rhob{ {\color{be} {\rho}} }
\def\varpib{ {\color{be} {\varpi}} }
\def\vb{{\color{be} v }}
\def\nablab{ {\color{be} \nabla}}
\def\Gammab{ {\color{be} \Gamma}}
\def\Thetab{ {\color{be} {\Theta}} }
\def\Ab{{\color{be} A }}
\def\Rb{{\color{be} R}}
\def\db{\spose {\raise 4.0pt \hbox{\color{be}{\,\bf-}}} {\rm d}}
\def\Sigmab{ {\color{be} {\mit\Sigma}} }
\def\Sb{ {\color{be} S } }
\begin{center}
{\color{rd}\bf  FIELDS IN NONAFFINE BUNDLES. II.\\[0.4cm]
Gauge coupled generalization of harmonic mappings \\
and their  Bunting identites.}
\\[1cm]
 \underline{Brandon Carter} \\[0.6cm]
 \textcolor{ru}{Group d'Astrophysique Relativiste (CNRS),
  Observatoire Paris - Meudon. }
  \\[0.5cm]
 {\color{be} 7 August, 1985.}
\\ [0.5 cm] [Colored version of article in {\it Phys.Rev.}
{\bf D33} (1986) 991-996].
\\[1.2cm]
\end{center}  .

{\bf Abstract. }  The general purpose bitensorially gauge-covariant 
differentiation procedure set up in the preceding article is specialised 
to the particular case of bundles with nonlinear fibres that are 
endowed with a torsion free Riemannian or pseudo-Riemannian structure.
This formalism is used to generalize the class of harmonic mappings between
Riemannian or pseudo-Riemannian spaces to a natural gauge coupled extension
in the form of a class of field sections of a bundle having the original
image space as fibre, with a nonintegrable gauge connection $\Amr$
belonging to the algebra of the isometry group of the fibre space. The
Bunting identity that can be used for establishing uniqueness in the
strictly positive Riemannian case with negative image space curvature
is shown to be generalizable to this gauge coupled extension.

\section{Introduction} 
\label{Section1}

The purpose of this article is to construct and investigate the natural
gauge coupled generalisation of the extensive class of non linear
field models known as harmonic mappings. The work will be based on the
use of the general-purpose bitensorially gauge covariant differentiation
formalism set up by the author in the preceding article \cite{1}, of which 
the relevant essentials (as specialised to the torsion free Riemannian or 
pseudo-Riemannian case) are summarised in Sec. \ref{Section2}  of the 
present article.

The general case of harmonic mappings has been studied in mathematical 
circles for many years. A convenient introductory review, from a 
physically motivated point of view, has been provided by Misner\cite{2}. 
One of the reasons for interest in harmonic mappings
{\be \cal M \mapsto \calXr \label{1.1}\fe}
is that they include the subclass known to physicists as nonlinear 
$\sigma$ models in the case when the Riemannian image space ${\calXr}$
has a suitable homogeneous symmetric space structure. Whenever ${\calXr}$ 
is subject to a continuous isometry group action (not necessarily a
fully effective one as in the homogeneous case) the possibility arises
of generalizing the class of simple mappings of the form (\ref{1.1}) to
bundle sections, whereby the base space ${\cal M}$ is mapped vertically
into fibres of the form ${\calXr}$ in a bundle ${\calBr}$ subject to
the isometry group in question. The purpose of the present work is to
describe the natural extension of the general class of harmonic mappings
of the form (\ref{1.1}) to a class of gauge-harmonic bundle sections
{\be {\cal M} \mapsto {\calBr} \label{1.2}\fe}
that will be automatically determined by the specification of a bundle 
connection $\Amr$ for any given Riemannian structure on the base 
${\cal M}$ and the fibre space ${\calXr}$. (For the special case of 
nonlinear $\sigma$ models with homogeneous symmetric space structure 
such a generalisation has already been carried out by the present 
author \cite{3} using the more traditional method whereby the curved 
fibre space is treated by an imbedding in a higher dimensional flat space.)

After the appropriate gauge harmonic field equations have been derived
in Sec. \ref{Section3}, they will be shown in Sec. \ref{Section4}
 to be amenable to treatment by an extension of the method recently 
developed in the context of ordinary harmonic mappings by 
Bunting\cite{4,5} for the purpose of establishing uniqueness of solutions 
subject to suitable boundary
conditions and inequalities.

\section{The bitensorially covariant differentiation procedure
for fibres with (pseudo-) Riemannian structure}
\label{Section2}

We start by summarizing the bitensorially covariant differentiation
procedure set up in the preceding article \cite{1} insofar as it 
applies to the restricted special case of fields taking values in a
space ${\calXr}$ with a Riemannian or pseudo-Riemannian
{\it metric structure} as specified in terms of local coordinates
$\Xr^{\!_\Ar}$ ( {\color{rd}\srm A} =1, ..., m) on ${\calXr}$ by
{\be {\rm d}\hat\srd^2= \hat \grd_{_{\Ar\Br}}\, {\rm d}\Xr^{\!_\Ar}\, 
{\rm d}\Xr^{\!_\Br} \label{2.1}\fe} 
with an associated metric connection whose components are given by the
standard formula
{\be \hat\Gammar{^{\ _\Br}_{\!_\Ar\ _\Cr}}=\hat \grd{^{_{\Br\Dr}}}
(\hat \grd_{_\Dr(_\Ar,_\Cr)} -\frac{_1}{^2}\hat \grd_{_{\Ar\Cr},_\Dr})
\label{2.2}\fe}
(using parentheses to denote symmetrization) where $\hat\grd{^{_{\Br\Dr}}}$ 
are components of the inverse metric to $\hat \grd_{_{\Ar\Br}}$ and the
comma suffixes indicate partial derivatives with respect to the 
corresponding coordinates.

Despite these restrictions in relation to the more general situation 
considered in the preceding article \cite{1} (where the connection
$\hat\Gammar$ was allowed to be quite arbitrary) the present context
remains nevertheless more general than that considered by Misner
\cite{2} insomuch as we do not suppose that the field configurations
in ${\calXr}$ are defined absolutely, but allow for the possibility
of an intrinsic indeterminacy modulo the action of a gauge group
${\calGr}$ which must, of course, be a subgroup of the isomorphism 
group of ${\calXr}$, the existence of a non-trivial gauge freedom thus
requiring the existence of a nontrivial isometry group with respect to
the metric (\ref{2.1}). This means that the field is to be considered
as a {\it section} $\Phir\{x\}$ in a bundle ${\calBr}$ with fibre
space ${\calXr}$ over a base space ${\cal M}$ which we shall suppose 
to be described by local ccordinates $x·^\mu$ ($\mu= 1, ... , n)$.
Such a bundle will be characterised locally by a simple but nonunique
direct-product structure which may be represented by expressing 
elements of a neighborhood in ${\calBr}$ as a couple $\{\Xr,x\}$ with 
corresponding local coordinates $\{\Xr^{\!_\Ar}, x^\mu\}$ for $\Xr\in
{\calXr}$, $x\in {\cal M}$. In such a coordinate system the
fibre-coordinate components of the fibre metric as induced on the
bundle will depend only on the $\Xr^{\!_\Ar}$, i.e. we shall have
{\be \hat \grd_{_{\Ar\Br}\, , \,\mu}=0 \label{2.3}\fe}
while, if we suppose that the base space has its own metric, given by 
{\be {\rm d}\sbe^2= \gbe_{\mu\nu}\, {\rm d}x^\mu\, {\rm d}x^\nu \fe} 
with the resulting connection
{\be \Gammab_{\!\mu\ \rho}^{\ \nu}=\gbe^{\nu\sigma}
(\gbe_{\sigma(\mu\, ,\,\rho)}-\frac{_1}{^2}\gbe_{\mu\rho\, ,\,\sigma})
\, ,\label{2.4}\fe}
the correponding form induced on the bundle will satisfy
{\be \gbe_{\mu\nu\, ,\, _\Ar}=0\, .\label{2.5}\fe}
The gauge indeterminacy consists in the possibility of conserving this
structure when effecting a fibre-coordinate transformation of the form
{\be \Xr^{\!_\Ar}\{\Xr,x\}\mapsto \Gru^{_\Ar}\{\Xr,x\}\, .\label{2.6}\fe}
The requirement that the property (\ref{2.3}) should be preserved is
expressible as the condition that the gauge transformation (\ref{2.6})
should be characterised by the property
{\be \Gru^{_\Ar}_{\ ,\, \mu}= \Gru^{_\Ar}_{\ ,\, _B} 
\hat \kru{^{_\Br}_{\ \mu}} \, ,\label{2.7}\fe}
where the $\hat\kru{^{_\Br}_{\ \mu}}$ are the components of some 
base-space one-form valued vertical vector field satisfying the fibre-space
Killing equation
{\be \hat\nablar{^{(_\Ar}}\hat\kru{^{_\Br)}_{\ \mu}}=0\, ,\label{2.8}\fe}
where $\hat\nablar$ is the ordinary Riemannian operation of
covariant derivation with respect to (\ref{2.1}) and (\ref{2.2}) as
expressed by
{\be \hat\nablar{^{_\Ar}} \hat\kru{^{_\Br}_{\ \mu}}= \hat \grd{^{_{\Ar\Cr}}}
(\hat\kru{^{_\Br}_{\ \mu\, , \, _\Cr}}+\hat\Gammar{^{\ _\Br}_{\!_\Cr \ _\Dr}} 
\hat\kru{^{_\Dr}_{\ \mu})} \, .\label{2.9}\fe} 

In such a context the gauge connection $\Amr$ can be appropriately 
envisaged, in the manner described in the preceding article \cite{1}, as 
a gauge-patch dependent fibre-tangent-vector-valued one-form with local 
coordinates $\Aru_\mu^{\, _\Ar}$ which acts as the generator of the relevant 
fibre-space isometry transformations, and which must therefor be 
characterised by the same Killing-vector property as
 $\hat \kru{^{_\Ar}_{\ \mu}}$,  namely
{\be \hat\nablar{^{(_\Ar}} \Aru_\mu^{\, _\Br)}=0\, .\label{2.10}\fe}
The condition characterised by (2.7) and (2.8) automatically ensures that
the corresponding gauge transformation
{\be \Aru_\mu^{\, _\Ar} \mapsto \Gru^{_\Ar}_{\ ,\, _\Br}
\Aru_\mu^{\, _\Br}- \Gru^{_\Ar}_{\, ,\mu} \label{2.11}\fe}
of the connector field $\Amr$ will preserve the gauge operator property
(\ref{2.10}). As in the more general context considered in the preceding
article \cite{1}, the nontensorial (inhomogeneous) transformation rule
(\ref{2.11}) gives rise to a purely tensorial transformation rule
$$ \Fru_{\mu\nu}^{\ \ _\Ar}\mapsto \Gru^{_\Ar}_{\ ,\, _\Br} 
\Fru_{\mu\nu}^{\ \ _\Br} $$ 
for the corresponding {\it fibre-tangent-vector-field-valued 
gauge-curvature two-form} $\Fmr$ as defined by the basic formula
{\be  \Fru_{\mu\nu}^{\ \ _\Ar}= 2 \Aru_{[\nu \ ,\mu]}^{\,\ _\Ar}+ 
2 \Aru_{[\nu}^{\ _\Br} \Aru_{\mu] \, ,_\Br}^{\ _\Ar} \label{2.12}\fe}
where square brackets denote antisymmetrization) so that ${\Fmr}$
can be considered as a {\it globally} well-defined bitensorial
field over the entire bundle ${\calBr}$

The covariant derivative $\Dru\Phir$ of a field $\Phir\{x\}$
(i.e. a section of the bundle ${\calBr}$) over ${\cal M}$ was shown
\cite{1} to have bitensorial components $\Phir^{_\Ar}_{\ \wru\, \mu}$
given simply by
{\be \Phir^{_\Ar}_{\ \wru\, \mu}=\partial_\mu \Xr^{\!_\Ar}+
\Aru_\mu^{\ _\Ar} \label{2.13}\fe}
and transforming under (\ref{2.6}) according to the ordinary vectorial rule
{\be  \Phir^{_\Ar}_{\ \wru\, \mu} \mapsto \Gru^{_\Ar}_{\ ,\, _\Br} 
\Phir^{_\Br}_{\ \wru\, \mu} \, , \label{2.14} \fe}
where $\partial_\mu \Xr^{\!_\Ar}$ denotes the base-space gradient components
of the coordinate components $\Xr^{\!_\Ar}\left\{\Phir\{x\}\right\}$ of the
field $\Phir$ with respect to the local gauge coordinate patch
$\{\Xr^{\!_\Ar}, x^\mu\}$ on the bundle ${\calBr}$.

In order to construct higher-order similarly bitensorial derivatives,
it is necessary to introduce the section dependent connector field
$\omega$ which was shown \cite{1} to be given by
{\be \omegaru_{\mu \ _\Br}^{\ _\Ar}=\Phir^{_\Cr}_{\ \wru\,\mu}\, 
\hat\Gammar{_{\! _\Br\ _\Cr}^{\, _\Ar}}
+ \Aru_{\mu \ ,\, _\Br}^{\ _\Ar}\, ,\label{2.15}\fe}
where the values of $\hat\Gammar{_{\! _\Br\ _\Cr}^{\  _\Ar}}$ and 
$\Aru_{\mu \,\ ,\, _\Br}^{\,\ _\Ar}$ are evaluated on the section $\Phir\{x\}$. 
In terms of this connector field, and of the ordinary base space 
connection $\Gammab$, the second-order covariant derivative components are 
expressible as
{\be \Phir^{_\Ar}_{\ \wru\, \mu\, \wru\, \nu}=
(\Phir^{_\Ar}_{\ \wru\, \mu})_{;\nu}+ \omegaru_{\nu\ _\Br}^{\ _\Ar}
\Phir^{_\Br}_{\ \wru\, \mu} \, ,\label{2.16}\fe}
where the semicolon denotes ordinary (base but not fibre) covariant
derivation as defined by
{\be (\Phir^{_\Ar}_{\ \wru\,\nu})_{;\mu}=
\partial_\mu\Phir^{_\Ar}_{\ \wru\,\nu}-\Gammab_{\!\mu \ \nu}^{\  \rho}
\Phir^{_\Ar}_{\ \wru\, \rho}\, .\label{2.17}\fe}
The antisymmetric part of this bitensor will be expressible (in the
torsion-free case under consideration here) purely in terms of the 
curvature field ${\Fmr}$ as evaluated on the section in the form
{\be \Phir^{_\Ar}_{\ \wru\,[\mu\, \wru\,\nu]}=\frac{_1}{^2}
\Fru_{\mu\nu}^{\ \ _\Ar} \, .\label{2.18}\fe}
Antisymmetrised differentiation at higher orders introduces contributions
arising from the curvature of the fibre and base spaces, as represented by 
the corresponding Rieman tensors with components 
$\hat \Rr{_{_{\Ar\Br} \ _\Dr}^{\,\ \ _\Cr}}$ and 
$\Rb_{\mu\nu \ \sigma}^{\,\ \ \rho}$ obtained 
from (\ref{2.2}) and (\ref{2.4}) by the standard formulae
{\be \hat \Rr{_{_{\Ar\Br} \ _\Dr}^{\,\ \ _\Cr}}= 2\, \hat
\Gammar{_{\![ _\Br\ |_\Dr|,_\Ar]}^{\, \ _\Cr}}+2\,\hat
\Gammar{_{\! [_\Ar\ |_\Er|}^{\, \ _\Cr}}
\hat\Gammar{_{\!_\Br]\ _\Dr}^{\, \ _\Er}}
\label{2.19a}\fe}
and 
{\be \Rb_{\mu\nu \ \sigma}^{\,\ \ \rho}= 2\,
\Gammab_{\![\nu\, \ |\sigma|, \mu]}^{\ \ \rho} + 2\,
\Gammab_{\![\mu\, \ |\tau|}^{\ \ \rho} \Gammab_{\!\nu]\ \sigma}
^{\, \ \tau}\, .\label{(2.19b}\fe}
Thus starting from the basic expression
{\be \Phir^{_\Ar}_{\ \wru\, \mu\, \wru\, \nu\, \wru\, \rho}=
(\Phir^{_\Ar}_{\ \wru\, \mu\, \wru\,\nu})_{;\rho}+ 
\omegaru_{\mu \ _\Ar}^{\ _\Br}
\Phir^{_\Ar}_{\ \wru\, \mu\, \wru\, \nu} \label{2.20}\fe}
for the third order bitensorially covariant derivative of $\Phir$,
one obtains 
{\be \Phir^{_\Ar}_{\ \wru\, \mu\, \wru\, [\nu\wr\rho]}=
\Omegaru_{\nu\rho\ _\Br}^{\ \ _\Ar} \Phir^{_\Br}_{\  \wru\, \mu} -
\Rb_{\nu\rho \ \mu}^{\,\ \ \sigma}\Phir^{_\Ar}_{\,\wru\,\sigma}
\, ,\label{2.21}\fe}
where the section dependent total curvature bitensor $\Omegaru$ will
be given \cite{1} (in this torsion-free case) by 
{\be \Omegaru_{\mu\nu\ _\Br}^{\ \ _\Ar}= \Phir^{_\Cr}_{\ \wru\,\mu}
\Phir^{_\Dr}_{\ \wru\, \nu} \hat \Rr{_{_{\Cr\Dr} \ _\Br}^{\,\ \ _\Ar}}+ 
\hat\nablar_{\!_\Br}  \Fru_{\mu\nu}^{\ \ _\Ar} \, .\label {2.22} \fe}

\section{The gauge coupled generalisation of a harmonic mapping}
\label{Section3}

In terms of the formalism set up in the preceding section, it is obvious 
how one should proceed to generalize the concept of a harmonic mapping
as described e.g. by Misner \cite{2}, so as to incorporate a minimal
gauge invariant coupling to a non-integrable gauge connector field.
In the following analysis we shall allow, in addition to the minimal gauge 
coupling, the possibility that there is also a non-linear gauge-invariant
nondifferential self-coupling field.

The field equations for such a system will be obtained by the application
of the usual kind of stationary-variation principle to a Lagrangian
integral of the form
{\be \calIr=\int {\rm d}^n x\, \Vert \gbe\Vert^{1/2} \Lr\{\Phir, \Dru\Phir\}
\label{3.1}\fe}
over the base space ${\cal M}$, where the Lagrangian scalar function
$\Lr$ is taken to be a quadratic function of the gradients of the
field section $\Phir$, with the gauge invariant form
{\be \Lr=\frac{_1}{^2}\rhob\, \hat \grd_{_{\Ar\Br}} \, \gbe^{\mu\nu}\,
\Phir^{_\Ar}_{\ \wru\, \mu}\Phir^{_\Br}_{\  \wru\,\nu} + \varpib
\hat\calVr\{\Phir\}\, ,\label{3.2}\fe}
where $\rhob$ and $\varpib$ are given scalar fields over the base
space ${\cal M}$ (which may occur naturally as known weight functions
in certain contexts) and where $\hat\calVr$ is a self-interaction
potential that is given as a scalar field over the fibre space $\calXr$
and which, to avoid breaking the syùùetry, should be required to be
invariant under the gauge group action, i.e. 
{\be \hat\kru{^{_\Ar}}\hat\calVr_{ ,\,_\Ar}= 0 \label{3.3}\fe}
for any member $\hat\kru{^{_\Ar}}$ of the subset of solutions of the 
fibre-space Killing equations
{\be \hat \nablar{^{(_\Ar}} \hat\kru{^{_\Br)}}=0 \label{3.4} \fe}
that constitute the gauge group algebra. (Evidently if one were 
considering the gauge coupling of the usual kind of nonlinear $\sigma$
model for which the fibre space $\calXr$ is homogeneous, and if one
wished to use a {\it maximal} gauge group which would act effectively
over the whole of $\calXr$, then the requirement (\ref{3.4}) would
restrict $\hat\calVr$ to be a trivial uniform field over $\calXr$ giving
no contribution to the field equations for $\Phir$.

The variation of the section $\Phir\{x\}$ and thus of the corresponding
coordinate components $\Xr^{\!_\Ar}\{\Phir\}$, while keeping the background 
fields $\rhob$, $\varpib$ as well as the base metric $\gbe$ and the gauge 
connection $\Amr$ constant, leads to an infinitesimal variation 
$\delta \Lr$ given in terms of the section component variations 
$\delta \Xr^{\!_\Ar}$ by
{\be \delta \Lr=(\rhob\,\Phir_{_\Ar}^{\ \wru\,\mu}\,\delta 
\Xr^{\!_\Ar})_{;\mu}+\frac{\delta \Lr} {\delta\Phir^{_\Ar}}\, 
\delta \Xr^{\!_\Ar}\, ,\label{3.5}\fe}
where the Eulerian derivative takes the form
{\be \frac{\delta \Lr} {\delta\Phir^{_\Ar}}=-\left[(\rhob\,
\Phir_{_\Ar}^{\ \wru\, \mu})_{\wru\,\mu} -\varpib \calVr_{, _\Ar}\right]
\ ,\label{3.6} \fe}
where the base and fibre metrics $\gbe$ and $\hat \grd$ have been used in 
the normal way for the definition of (bitensorially) covariant index 
raising and lowering, so that explicitly
{\be \Phir_{_\Ar}^{\ \wru\,\mu}=\gbe^{\mu\nu}\, \hat \grd_{\Ar\Br}
(\partial_\nu \Xr^{\!_\Br} + \Aru_\nu^{\  _\Br})\, .\label{3.7}\fe}
In expressing the first term on the right hand side of (3.5), use has
been made of the fact that because it acts on a quantity that is scalar 
with respect to the fibre indices, the ordinary base-coordinate
covariant differentiation opration, as denoted by a semicolon, can be
used interchangeably with the gauge covariant differentiation operation
indicated by a heavy bar. Since it thus takes the form of an ordinary 
divergence this first term can be eliminated in the usual way, so that 
one obtains the required field equations 
{\be \frac{\delta \Lr} {\delta\Phir^{_\Ar}}=0 \label{3.8}\fe}
expressing the condition that $\Phir$ should be a critical point of the
integral $\calIr$, in the form
{\be (\rhob\,\Phir_{_\Ar}^{\ \wru\, \mu})_{\wru\,\mu} =\varpib\hat
\calVr_{, _\Ar}\  .\label{3.9} \fe}
This system of equations may be written out in somewhat more explicit
but no longer manifestly gauge dependent form as
 {\be \nablab^\mu(\rhob\,\Phir^{_\Ar}_{\ \wru\,\mu})+\rhob\,
\Phir^{_\Br\,\wru\,\mu}\hat \nablar_{\!_\Br} \Aru_\mu^{\ _\Ar}=
\varpib \hat \nablar^{_\Ar}\hat\calVr\, ,\label{3.10}\fe}
where (using the notation of the preceding article \cite{1}) 
{\be \nablab^\mu(\rhob\,\Phir^{_\Ar}_{\ \wru\,\mu})=(\rhob\,
\Phir^{_\Ar}_{\ \wru\,\mu})^{;\mu}+\rhob\, \Phir^{_\Br\,\wru\,\mu}\hat
\Gammar{_{\!_\Cr\ _\Br}^{\ _\Ar}}\partial_\mu \Xr^{_\Cr}\, .\label{3.11}\fe}
In transforming from (\ref{3.9}) to (\ref{3.10}) we have taken advantage of
the fact that, in addition to the obvious consequence
{\be \gbe_{\mu\nu\, \wru\, \rho}=0 \label{3.12}\fe}
of (\ref{2.4}), the conditions (\ref{2.2}),  (\ref{2.3}), and (\ref{2.10})
taken together ensure that the gauge covariant derivative of the fibre
metric $\hat \grd$, as evaluated on the section $\Phir$, will also
automatically vanish, i.e.
{\be \hat \grd_{_{\Ar\Br}\,\wru\,\mu}=0 \label{3.14} \fe}
so that all the bitensorial index raising and lowering operations
commute with bitensorially covariant differentiation.

In the absence of the additional self-coupling term $\hat\calVr$ and of the
gauge field $\Amr$ (and provided $\rhob$ is uniform) the field equations
(\ref{3.10}) can be seen to reduce to a system of the much studied
harmonic type described e.g. by Misner \cite{2}.

\section{Generalised Bunting identity for gauge harmonic mappings}
\label{Section4}

The purpose of this final section is to extend to the gauge-coupling
system that has just been presented a very useful identity, involving the
deviation between two hypothetically different sections, of the kind that
was introduced by Bunting \cite{4,5} for sytems of ordinary harmonic type
for the purpose of establishing uniqueness subject to appropriate 
inequalities and boundary conditions. Bunting's work was motivated by the
problem of establishing the uniqueness of solutions of the black hole
equilibrium problem, which had been reduced by the present author 
\cite{6,7} (subject to global hypotheses which still lack an entirely 
complete and rigourous justification) to a boundary value problem of the 
right (harmonic) type. A concise presentation of the main results of 
Bunting's work, and an examination of the relationship between the Bunting 
identity and a more specialized identity constructed for the same purpose 
by Mazur \cite{8} (including as a special case the identity of Robinson 
\cite{9}) has recently been given by the present author \cite{10}.
The Mazur identity applies only in the more restricted context (which, 
however, includes the case of the black hole problem) in which the
harmonic system is an appropriate kind of nonlinear $\sigma$ model, as
characterised by a requirement to the effect that the image (fibre) space 
should have a fully symmetric homogeneous structure. The natural gauge 
coupled extension of the Mazur identity for such fully symmetric spaces has
already been described by the present author elsewhere \cite{3}. The
Bunting method, which is applicable to a much less restricted class of 
image spaces, uses concepts introduced in  general study of harmonic 
systems by Schoen and Yau \cite{11}.

The context that we wish to consider is one in which we have two distinct
bundle sections $\Phir_{[0]}\{x\}$ and  $\Phir_{[1]}\{x\}$ which we suppose
to be homotopically connectable in the sense that in the fibre over each
base point $x\in {\cal M}$ there is some (smooth) curve $\Phir\{t,x\}$
parametrised by a variable $t$ ranging from 0 to 1, with
{\be \Phir\{0;x\} =\Phir_{[0]}\{x\}\, ,\hskip 1 cm
 \Phir\{1;x\} =\Phir_{[]1}\{x\} \label{4.1}\fe}
and varying smoothly as a function of $x$ so that $\Phir\{t,x\}$
represents a well behaved section in $\calBr$ over ${\cal M}$ for each 
fixed value of $t$. Without loss of generality -- except for the exclusion
of curves that become null in fibre spaces with indefinite metric signature
-- we may, following Bunting, require that the parametrization should be
adjusted so as to be affine along the curve above each base point
$x\in {\cal M}$, with parametrization chosen so that the corresponding
tangent vector with components given by
{\be \hat \srd{^{_\Ar}}=\frac {{\rm d}\Xr^{_\Ar}}{{\rm d} t} \label{4.2}\fe}
should everywhere satisfy
$$ \hat \srd^{_\Ar} \hat \srd_{_\Ar}=\hat\srd^2 \, ,$$
where $\hat \srd$ is the total metric length of the curve.

Let us introduce the notation
{\be \hat\Dr=\hat \srd{^{_\Ar}}\hat\nablar_{\!\Ar} \label{4.3}\fe}
to denote the correspondingly parametrized operation of covariant
differentiation along the fibre above any {\it fixed base point}
$x\in {\cal M}$. Let us similarly use the symbol $\Dru_\mu$ as defined
in Sec. \ref{Section2}, of quantities defined in the section
$\Phir\{t,x\}$ defined by any {\it fixed affine parameter value} $t$.
When applied successively to the fields $\Phir\{t,x\}$ in $\calBr$,
these gauge covariant differentiation operators satisfy
{\be\hat\Dr\,(\rhob\, \Dru_\mu\Phir^{_\Ar})-\rhob\, 
\Dru_\mu\hat\srd{^{_\Ar}}= 0 \, . \label{4.4}\fe}
At the next higher order, they satisfy a commutation identity that
involves the fibre space curvature, in the form
{\be \hat\Dr\big( \Dru_\nu(\rhob\, \Dru_\mu\Phir^{_\Ar})\big)
-\Dru_\nu(\rhob \Dru_\mu\hat\srd^{_\Ar})=\rhob\, \hat\Rr
{^{_\Ar}_{\ _{\Br\Cr\Dr}}}(\Dru_\mu\Phir^{_\Br})\,\hat\srd{^{_\Cr}}
\Dru_\nu\Phir^{_\Dr}\, .\label{4.5}\fe}

Now the squared path length $\hat\srd^2$ appearing in (\ref{4.2}) can be
considered as an ordinary (evidently gauge independent) scalar field
over the base ${\cal M}$. As such it will have gradient components given by
{\be \frac{_1}{^2} \rhob\, (\hat \srd^2)_{;\mu}=\frac{_1}{^2} \rhob\,
\Dru_\mu (\hat \srd^2)=\rhob\, \hat\srd_{_\Ar}\Dru_\mu\hat\srd{^{_\Ar}}
\, ,\label{4.6}\fe}
where the right hand side is to be evaluated at any fixed value of $t$
in the interval $\{0,1\}$, the result being manifestly independent of
the choice. Taking the base-space divergence of this relation gives
{\be \frac{_1}{^2}\big( \rhob\, ( \hat \srd^2)_{;\mu}\big)^{;\mu}=\rhob\,
(\Dru^\mu\hat\srd_{_\Ar})\Dru_\mu\hat\srd{^{_\Ar}}+\hat\srd_{_\Ar}
\Dru^\mu(\rhob\, \Dru_\mu\hat\srd{^{_\Ar}})\, .\label{4.7}\fe}
Now since the left-hand side is manifestly independent of $t$, the same
must hold for the apparently $t$ dependent right-hand side, which will
therefore be unaffected by integration with respect to $t$ over the
unit interval $\{0,1\}$. Applying this integral operation to the second
term on the right, and using (\ref{4.5}), one obtains, via integration 
by parts,
$$ \int_0^1\!\! {\rm d}t\,\big(\hat\srd_{_\Ar}\Dru^\mu(\rhob\,\Dru_\mu
\hat\srd{^{_\Ar}})\big)= \big[\hat\srd_{_\Ar}\Dru^\mu(\rhob\, \Dru_\mu
\Phir^{_\Ar})\big]_0^1 \hskip 7.6 cm $$
{\be \hskip 4.2 cm -\!\int_0^1\!\!\! {\rm d}t \big((\hat\Dr\hat\srd_{_\Ar})
\Dru^\mu(\rhob\, \Dru_\mu\Phir^{_\Ar})+\rhob\,\hat\srd_{_\Ar} \hat\Rr
{^{_\Ar}_{\ _{\Br\Cr\Dr}}}(\Dru_\mu\Phir^{_\Br})\,\hat\srd{^{_\Cr}}
\Dru_\nu\Phir^{_\Dr}\big).\label{4.8}\fe}
It is to be noticed that the end-point contributions in the first term on the
right hand side of this relation are proportional (via a contraction with
$\hat\srd_{_\Ar}$) to the first term in the field equation (\ref{3.9}).
We can construct an analogous expression involving the other term in
(\ref{3.9}) by a similar integration by parts of the form
{\be  \int_0^1\!\! {\rm d}t\,\big(\hat\srd{^{_\Ar}}\hat\srd{^{_\Br}}
\hat\calMr_{_{\Ar\Br}}\big)=\big[\hat\srd{^{_\Ar}}\hat\calVr_{,_\Ar}\big]_0^1
-\!\int_0^1\!\!\! {\rm d}t\, 
\big((\hat\Dr\hat\srd{^{_\Ar}})\hat\calVr_{,_\Ar}\big)\, ,\label{4.9}\fe}
where the symmetric mass tensor field on the fibres is defined by
{\be \hat\calMr_{_{\Ar\Br}}=\hat\nablar_{\!_\Ar}\hat\nablar_{\!_\Br}
\hat\calVr \, .\label{4.10}\fe}
Combining (\ref{4.7}),  (\ref{4.8}), and (\ref{4.9}) we can obtain
an identity of the form
$$ \frac{_1}{^2}\big( \rhob\, ( \hat \srd^2)_{;\mu}\big)^{;\mu}+\left[
\frac{\delta\Lr}{\delta\Phir^{_\Ar}} \hat \srd{^{_\Ar}}\right]_0^1
-\int_0^1 \!\! {\rm d}t\,\frac{\delta\Lr}{\delta\Phir^{_\Ar}}\,
\hat\Dr\hat\srd{^{_\Ar}}=\hskip 6.6 cm$$
{\be\hskip 2.6 cm \int_0^1\!\! {\rm d}t\,\big(\rhob(\Dru^\mu
\hat\srd_{_\Ar})\Dru_\mu\hat \srd{^{_\Ar}}-\rhob\,\hat\Rr_{_{\Ar\Br\Cr\Dr}}
\hat\srd{^{_\Ar}}(\Dru^\mu\Phir^{_\Br})\hat\srd{^{_\Cr}}\Dru_\mu
\Phir^{_\Dr}+\varpib\, \hat\calMr_{_{\Ar\Br}}\hat\srd{^{_\Ar}}
\hat\srd{^{_\Br}}\big) \label{4.11}\fe}
with ${\delta\Lr}/{\delta\Phir^{_\Ar}}$ as given by (\ref{3.6}), so that
the left-hand side evidently resuces to a (weighted) Laplacian of the
squared fibre distance $\hat\srd^2\{x\}$ between the sections 
$\Phir_{[0]}\{x\}$ and $\Phir_{[1]}\{x\}$ at each base point $x\in{\cal M}$
if  {\it all} the intermediate sections $\Phir\{t;x\}$ of the homotopy
are solutions of the field equations (\ref{3.8}).

For the purpose of establishing uniqueness of solutions to the 
appropriate global boundary condition problems, it will {\it not} do to 
assume that the intermediate sections $\Phir\{t;x\}$ are solutions of the 
field equations: in examining the possible deviation -- as measured by 
$\hat\srd{^2}$ -- between the solutions  $\Phir_{[1]}\{x\}$ and
$\Phir_{[0]}\{x\}$ one will be justified in setting ${\delta\Lr}/
{\delta\Phir^{_\Ar}}$ equal to zero {\it only} at the end points
$t=0$ and $t=1$. We can however get rid of the integral involving the
values of ${\delta\Lr}/{\delta\Phir^{_\Ar}}$ at intermediate values of $t$
if we now {\it restrict} the homotopy (which up till this stage has been 
left arbitrary apart from the parametrization) by requiring that the
fibre space curves  $\Phir\{t;x\}$ for each {\it fixed} base point $x$
should be {\it geodesic}, which means (since the parametrization has 
already been restricted to be affine) that they should satisfy the 
equation
{\be \hat\Dr\hat\srd{^{_\Ar}}=0\, .\label{4.12}\fe}
In practice this is likely to be a less serious restriction than might at
first appear, since one of the most convincing ways of establishing the
existence of the homotopy itself will be to construct it explicitely in
terms of geodesics in the first place: all that is needed is to be sure 
that for each base point $x \in {\cal M}$ the required geodesic exists
and that it is unique, at least subject to conceivably relevant
additional restrictions (e.g. on the maximum allowed length of $\hat\srd$)
of such a nature as to guarantee the continuous variation of the geodesic 
as a function of the end points.

In the particular case of a fibre space with a complete {\it positive} 
definite metric and a  {\it negative} definite Riemannian curvature, 
then it is a well known theorem (see e.g. Kobyashi 
and Nomizu\cite{12}) that the required geodesic between any two points 
exists and is unique, thereby establishing the existence (and uniqueness)
of the required geodesic homotopy. Under these conditions, integration
of (\ref{4.11}) over a base domain $\Sigmab$ [using the same volume
measure as in the variational integral (\ref{3.1})] gives
{\be \oint_{\!\Sb=\partial\Sigmab}\!\!\!\!\!\!\!\!\!\! \rhob\,\hat\srd\,
\hat\srd{^{;\mu}}\, {\rm d}\Sb_\mu\!=\!\! \int_{\!\Sigmab}\!\!\! 
{\rm d}\Sigmab\!\!\int_0^1\!\!\!\!\! {\rm d}t\big( \rhob
(\!\Dru^\mu \hat\srd_{_\Ar})\Dru_\mu\hat\srd{^{_\Ar}}\!-\!\rhob\,\hat
\Rr_{_{\Ar\Br\Cr\Dr}}\hat\srd{^{_\Ar}}(\!\Dru^\mu\Phir^{_\Br})\hat\srd{^{_\Cr}}
\Dru_\mu \Phir^{_\Dr}\!+\!\varpib\, \hat\calMr_{_{\Ar\Br}}\hat\srd{^{_\Ar}}
\hat\srd{^{_\Br}}\big)\, , \label{4.13}\fe}
where the right hand side here will be a manifestly positive definite 
function of the distance $\hat\srd$ (vanishing only if $\hat\srd=0$, i.e. 
if the two solutions coincide) provided that the density $\rhob$ and also 
the base metric  $\gbe$ are positive definite, and that the potential 
$\hat\calVr$ has the appropriate convexity property as expressed by the 
condition that the mass tensor $\hat\calMr$ (weighted by $\varpib$)
should be positive definite. Under such conditions, it will suffice if
the boundary conditions ensure the vanishing of the surface integral
on the left-hand side of (\ref{4.13}) in order for one to be able to
conclude that $\hat\srd$ must vanish throughout the domain, and thus that
the solution $\hat\Phir$ is unique (for the given fibre metric $\hat\grd$,
connection $\Amr$, self-interaction potential $\hat\calVr$, and the given
base-space fields $\gbe$, $\rhob$, $\varpib$).

Even if the require negativity of the fibre-curvature tensor and positivity 
of the mass tensor held only indefinitely -- as might be the case, for 
example, if the potential $\hat\calVr$ were absent -- then the right-hand 
side of (\ref{4.13}) would still be positive definite as a function of 
$\Dru_\mu\hat \srd{^{_\Ar}}$. In these rather more general circumstances,
boundary conditions ensuring the vanishing of the surface integral on the
left hand side of (\ref{4.13}) would be sufficient, by (\ref{4.6}), to 
guarantee at least the vanishing of the gradient of $\hat\srd$, i.e.
$\hat \srd_{,\,\mu}=0$.
After one had thus established the uniformity of the fibre-metric distance 
between the two hypothetical solutions, it would suffice in addition for
the field to be fully determined even at just a single point on the
boundary $\Sb$, in order to be able to conclude that $\hat \srd$ vanishes
everywhere and thus that the solution is entirely unique

\end{document}